# Reviewer Assignment Problem: A Scoping Review

Jelena, J.J., Jovanovic

University of Belgrade, Serbia, jelena.jovanovic@fon.bg.ac.rs

Ebrahim, E.B., Bagheri

Toronto Metropolitan University, Canada, bagheri@ryerson.ca

The quality of peer review, and consequently the published research, depends to a large extent on the ability to recruit adequate reviewers for submitted papers. However, finding such reviewers is an increasingly difficult task. To alleviate this challenge, numerous solutions have been suggested for automated association of papers with "well matching" reviewers – the task often referred to as *reviewer assignment problem* (RAP). Yet, to our knowledge, a recent systematic synthesis of the RAP-related literature is missing. To fill this gap and support further RAP-related research, we present a scoping review on computational approaches to RAP. Following the latest methodological guidance for scoping reviews, we have collected and synthesised the literature on RAP published over the last five years (Jan 2016 - Sept 2021). Specifically, the review is focused on the following aspects of RAP research: i) the overall approach to RAP; ii) the criteria used for reviewer selection; iii) the modelling of candidate reviewers and submissions; iv) the computational methods for matching reviewers and submissions; and v) the evaluation methods used for assessing the performance of the proposed solutions. The paper summarises and discusses the findings for each of the aforementioned aspects of RAP research and suggests future research directions.

CCS CONCEPTS • General and reference → Surveys and overviews • Information systems → Expert search • Information systems → Recommender systems • Information systems → Decision support systems • Applied computing → Publishing

**Additional Keywords and Phrases:** reviewer assignment problem, reviewer recommendation, automated reviewer assignment, scoping review, peer review

## 1 INTRODUCTION

Peer review is quintessential for assuring the quality and continuous progress of scientific research. However, finding experts who would be able and willing to take on the role of reviewers for different scientific venues (e.g., journals, conferences, workshops) is not an easy task. In fact, according to the Global State of Peer Review report [40], journal editors report that finding willing reviewers is the hardest part of their job. Over time, this task is becoming even more difficult due to the continuous increase in the production of scientific papers and the consequent rising need for peer reviews [5]. This is further compounded by the fact that reviewer acceptance and completion rates have been steadily declining for several years now [46]. Also, review requests are often sent to a small subset of potential reviewers, which does not only limit the reviewer pool but also introduces a bias in the reviewer selection process [40].

Considering the above stated challenges associated with the task of finding suitable and willing reviewers, automated solutions have been researched as a potential way to alleviate these challenges and make the reviewer assignment task more efficient, faster, and less biased. Over the last three decades, since automated assignment of reviewers to papers has become a research topic - often referred to as *reviewer assignment problem* (RAP) - a variety of technical solutions have been proposed (see the Backgrounds section). However, it is unclear what the current state of development in this area is, since systematic reviews of the RAP-related literature are scarce. In particular, to our knowledge, the latest review of RAP solutions was published in 2010 [47], and even that review was not based on a systematic review process. We have identified this lack of a more recent review as an important gap in the literature, especially considering that over the last few years, technologies and computational methods relevant to RAP have made significant progress. Furthermore, for any research field, RAP included, to make further headway, it is important to shed light on the state of the art in the field and identify challenges lying ahead. This has been our main motivation to synthesise knowledge on recent RAP-related research.

This paper presents a scoping review of the literature on RAP, published over the last five years, that is, from January 2016 till September 2021. We have focused on the last five years since we intended to cover the period that roughly coincides with the significant developments in natural language processing (NLP) research (such as neural embeddings [28, 24] and advanced language models [41, 50]). In particular, since NLP methods and techniques have been an intrinsic part of a large majority of RAP approaches (to analyse submissions' content and candidate reviewers' publications), it would be expected that the progress in NLP would be reflected in RAP research. Furthermore, our preliminary review of the RAP-related literature suggested that over the last few years, the proposed solutions included reviewer selection / assignment criteria that 'traditional' approaches did not consider - e.g., research interests of candidate reviewers, the quality of their earlier reviews, or different ways of identifying potential conflicts of interest. To explore how these and other developments have shaped the RAP landscape, we have adopted the latest methodological guidance for conducting [35] and reporting [45] scoping reviews to guide us through collecting, synthesising, and presenting information on several aspects of RAP research including i) the overall framing of and approach to RAP; ii) the criteria used for reviewer selection; iii) the modelling of candidate reviewers and submissions; iv) the computational methods for matching reviewers and submissions; and v) the evaluation methods used for assessing the performance of the proposed solutions.

## 2 BACKGROUND

The quality and trustworthiness of scientific publications are often established by the critical evaluation of the methodology and reported findings, through a peer review process. However, finding appropriate reviewers for a particular submission (in case of a journal) or a set of submissions (in case of a conference or a workshop) is a demanding and time-consuming task [38]. The difficulties lie in the fact that candidate reviewers should not only have expertise in the topics of the submissions, but should also meet other criteria, such as, absence of conflict of interest and availability. Furthermore, since reviews are typically done on a voluntary basis, with no or very little explicit recognition for the completed work, one also needs to account for the interests of candidate reviewers, to increase the chance that an invited reviewer would accept the review invitation [49]. In addition, in each field, and especially in the rapidly expanding ones, there is a need for a continuous expansion of the reviewer pool, as a way of preventing the creation of research cliques and assure good coverage of emerging research topics [4].

Faced with these difficulties, the research community has developed different mechanisms aimed at facilitating the task of pairing submissions with appropriate reviewers. Paper bidding and keywords-based matching are among the most widespread ones [38], though each with its own shortcomings. For example, only a small proportion of potential reviewers are willing to engage with a bidding process and even when they do, they often get discouraged by the sheer volume of the submissions and thus only partially complete the bidding task. On the other hand, keywords-based matching methods are limited due to their reliance on pure syntactic matching, without the ability to deal with ambiguous and homonymous topics and subject areas. These and other challenges led to the emergence of advanced automated and semi-automated methods for assigning reviewers to submission, with the objective of making the reviewer assignment task more efficient, faster, and less biased. Since the seminal paper by Dumais and Nielsen [8], which proposed the use of latent semantic indexing [6] for automated topic-based matching of papers and reviewers, the research community has explored a variety of approaches to automate reviewer assignment. This research topic is often referred to as reviewer assignment problem (RAP), the term we use throughout the paper to refer to the task of automated assignment of reviewers to submissions.

Methods-wise, RAP research can be categorised into two broad categories: i) information retrieval (IR) based approaches and ii) optimisation-based approaches. IR-based methods are typically focused on devising new or adapting existing text representation methods (e.g., vector space model or topic modelling methods) to create profiles of submissions and candidate reviewers based on their content (e.g., title and abstract) and evidence of expertise (e.g., titles and abstracts of published papers), respectively. Data required for creating reviewer profiles are typically collected from publicly available academic data repositories (e.g., DBLP[1], AMiner[2] or Microsoft Academic Graph[3]) or provided directly by reviewers (e.g., reviewers may be asked to upload a selection of their own papers and/or other textual evidence of their expertise). The creation of reviewer and submission profiles is followed by estimating the level of matching between the submission and the candidate reviewer profiles (e.g., using a measure of similarity), sorting the candidates based on the estimated matching levels, and selecting top-N candidates to recommend as reviewers for the given submission.

The second group, optimization-based RAP approaches, also considers the level of matching between candidate reviewers' expertise and submissions. The maximisation of the overall matching (i.e., matching across all submissions) is often set as the objective function, which is optimised under a set of constraints that typically include absence of conflicts of interest (COI), reviewer workload limits (i.e., maximum number of papers assigned for review), and the required number of reviewers per submission.

Whereas IR-methods have typically been suggested for journal-like RAP scenarios (i.e., in cases where a journal editor looks for reviewers for a particular paper), optimization-based approaches are more often adopted for conference-like scenarios where reviewers need to be assigned to a large number of submissions in batches. In both groups of approaches, in addition to considering the expertise of candidate reviewers and their level of relevance to the submission's content, additional selection criteria have often been considered, including, but not limited to, authority, diversity, (topical) coverage, and availability of candidate reviewers. For example, Liu et al. [25] propose a graph-based approach that incorporates expertise, authority, and diversity (in topics of expertise) when ranking candidate reviewers for a particular submission. In another study, having recognised that research papers often cover more than one topical area, Karimzadehgan et al. [16] propose a method that considers multiplicity of paper topics and assigns a panel of reviewers who can collectively cover all topics of the paper. More recent RAP approaches consider research interests of candidate reviewers (as a way of increasing the likelihood of review request acceptance) by computing indicators of interest trend and including them in the optimisation function [13] or by putting more weight on topics of reviewers' recent publications in topic-wise matching of reviewers and submissions [37]. Furthermore, absence of COI is considered in both groups of approaches and while in some works it is treated as given (i.e., provided by reviewers), in others, it is derived from one or more types of author-reviewer relationships (such as co-authorship, co-workership, or advisor-advisee relationship) (e.g., [26, 22]).

---

[1] http://dblp.org
[2] https://www.aminer.org/
[3] https://www.microsoft.com/en-us/research/project/microsoft-academic-graph/

Many commercial conference management systems and journal editorial management systems offer more or less advanced support through automated reviewer assignment and / or recommendation. Examples include Elsevier's Find Reviewer Using Scopus[4], Springer Nature's Reviewer Finder[5], and Microsoft's Conference management toolkit[6]. The latter integrates the Toronto matching system [4], which is probably among the better-known examples of RAP research solutions that has been successfully applied in practice.

Even though a lot has been done and significant progress has been made in the automation of the reviewer assignment task, there is still room for improvement, as journal editors and conference program chairs are still struggling with finding sufficient numbers of competent and willing reviewers [49, 46]. To make further progress, it is important to have a good understanding of the current solutions, their framing of RAP, their underlying computational methods and techniques, and their performance. The scoping review reported in this paper aims to offer these insights.

## 3 RESEARCH OBJECTIVES AND QUESTIONS

Our overall objective is to provide a comprehensive insight into recent developments in RAP-related research, that is, computational approaches for automated assignment of reviewers to submissions. Accordingly, the primary research question (RQ) that guides our scoping review is defined as follows: *What computational approaches have been proposed and evaluated for solving the reviewer assignment problem (RAP)?* Following the guidance for conducting scoping reviews [35], we have also defined several specific RQs, each one focused on a specific aspect of RAP computational solutions:

RQ1: How has RAP been framed and what approaches have been applied to solve RAP?

RQ2: What criteria have been used for reviewer assignment / recommendation?

RQ3: What representation modelling approaches have been used for representing candidate reviewers and submissions?

RQ4: What computational methods have been used for assigning / recommending reviewers to submissions?

RQ5: How have RAP solutions been evaluated? What were the main features (datasets, benchmarks, evaluation measures) of the conducted evaluation studies?

## 4 METHODS

### 4.1 Study Selection Procedure

We conducted a systematic literature search using Scopus, Google Scholar, and DBLP databases. Search terms were chosen to reflect the review objectives, that is, to capture recent studies that proposed computational methods for addressing the RAP. Standard Boolean operators were used to combine the selected keywords into search queries. The search queries that were used in our work are listed in Table 1.

Table 1. Search queries

| |
|---|
| "reviewer assignment problem" |
| "reviewer recommendation" |
| ( reviewer AND assign* ) OR ( reviewer AND recommend* ) |

---

[4] https://service.elsevier.com/app/answers/detail/a_id/29385/supporthub/publishing/
[5] https://www.springernature.com/gp/editors/resources-tools/reviewer-finder
[6] https://cmt3.research.microsoft.com/About

To identify relevant studies in the search results, we defined specific inclusion and exclusion criteria. As per the guidance for scoping reviews [35], we defined these criteria in relation to the scoping review's main concept, its context, and sources of evidence. The concept-related criteria include those defining the overall topical focus of the review and the RAP-specific focus, the context-related criteria define the peer review context we focus on, whereas the criteria defining the sources of evidence include the date, type, and language of the publication, the recency and completeness of the reported approach, and the quality of the reported evaluation. Details of the eligibility criteria are provided in Table 2.

Table 2. Inclusion and exclusion criteria

| Criterion | | Inclusion | Exclusion |
| --- | --- | --- | --- |
| Source of evidence | Period | Jan 2016 - Sept 2021 | Outside these dates |
| | Language | Articles written in English | Articles in languages other than English |
| | Publication type | Journal and conference articles | Book chapters, abstracts, review papers |
| | Recency and completeness | In case of papers that report on the same or only slightly changed (improved) method, include only the paper presenting the latest or the most complete (if published in the same year) report on the method. | Papers reporting on an older or less complete version of a method that has been published later and/or in a more comprehensive manner (e.g., a conference paper presenting a method that was later published in a journal). |
| | Evaluation | Studies that evaluate the proposed approach through comparison with meaningful baselines on real-world datasets (or simulations of such datasets). Baselines, as defined here, include alternative RAP methods and / or "ground truth" (e.g., assignments by human experts). | Solutions evaluated on examples or toy datasets; evaluation studies that do not include comparison with baselines. |
| Concept | General topical focus | Automated assignment / recommendation of reviewers to submissions in academic peer review settings. | Code review (in software development); peer review (of student assignments) in educational settings; expert finding. |
| | RAP specific focus | Studies that propose and evaluate a new or improved automated or semi-automated approach to reviewer assignment / recommendation. | Studies with primary or exclusive focus on other aspects of peer review or reviewer assignment (e.g., fairness, bias) |
| Context | Peer review context | Typical academic peer review context (journals, conferences, workshops, majority of grant agencies).<br>It is assumed that reviewers express their opinion (evaluation) as the cardinal ranking of the reviewed submissions, as is the practice in a large majority of academic peer review contexts. | Review of grant proposals where granting agencies follow idiosyncratic review procedures; other specific peer-review contexts and procedures (e.g., procedures that require reviewers to do pairwise comparisons of the assigned submissions and express their opinion in the form of ordered ranking of the reviewed submissions) |

## 4.2 Results Synthesis Procedure

To extract and summarize relevant data from the selected articles, we define several groups of data charting variables as explained in Table 3. The definition of these variables was guided by the review objectives and RQs, and they are related to i) the data used for representing (modelling) candidate reviewers (CRs) and submissions; ii) the overall framing of the RAP task; iii) the criteria for reviewer assignment / recommendation; iv) the computational methods for modelling of CRs and submissions, estimating the level of matching between them, and assigning / recommending reviewers to submissions; and v) the evaluation method.

Table 3. Data charting variables

| Variable group | Variable | Description |
| --- | --- | --- |
| Data used for representing candidate reviewers (CRs) and submissions | Collection of data about CRs | Was data collection about CRs a part of the proposed solution? If it was, in what ways the data were collected and what sources were used for data collection? |
| | Kinds of data used for modelling CRs and submissions | For example, for submissions: title and abstract only, or also the authors, keywords, and the like; for CRs: only their publications (titles and abstracts), or also authority metrics (such as h-index) and/or some other data |
| Overall framing of the RAP task | The overall computational approach to RAP | Was the proposed approach framed as, for example, a recommendation task, a classification task, a constrained-based optimization task, or in some other way? |
| | Individual or group focus | Was the objective to recommend individual reviewers or groups of reviewers? |
| Criteria for reviewer assignment / recommendation | | What criteria were used for reviewer assignment / recommendation? This could include, for example, expertise, interests, conflict of interest, etc. If a group was to be recommended, was the complementarity and/or diversity of group members used a criterion? |
| Computational methods | Methods for representing (modelling) of CRs and submissions | This could include, for example, topic models, author topic models, network models, etc. |
| | Methods for estimating the level of matching between a submission and a CR | This could be, for example, Latent Semantic Indexing or some similarity measures |
| | Methods for assigning / recommending reviewer(s) to a submission | This could, for example, take the form of returning the top K matching CRs or a sophisticated assignment algorithm |
| Evaluation method | The overall evaluation approach | A summary of the evaluation study setup |
| | Datasets used | The characteristics of the data set used for evaluating the proposed method |
| | Baselines | The kinds of baseline used, such as RAP methods reported in earlier work; a general modelling or optimisation method; the "ground truth" |
| | Evaluation measures | Different kinds of measures used for comparing the proposed method with the baselines |
| | Replication readiness | The availability of i) the code that implements the proposed method, ii) the used datasets, iii) the experimental setup |

## 5 RESULTS

### 5.1 The Selection Evidence Sources

The intermediate and final results of the study selection process are given in Figure 1. Our search queries generated 1,808 search results across the three selected academic databases. Based on the initial (title-based) screening of the search results, we excluded 1,315 articles. After importing the remaining articles into a spreadsheet (Google Sheets) for further analysis, 132 duplicates were detected and removed. The remaining 362 articles were screened based on their abstract and keywords.

As a result, 315 articles were excluded, whereas the remaining ones (N=47) were used as the input for backward snowballing, that is, their references were examined to identify additional potentially relevant articles that our initial search missed. This led to the detection of four additional potentially relevant articles, which means that 51 articles were eventually selected for the full-text screening based on the inclusion / exclusion criteria listed in Table 2. The result of this final step was the selection of 26 articles for the review. The overall study selection process was performed by two reviewers (paper authors), independently; a few disagreements that occurred along the way were resolved through discussion.

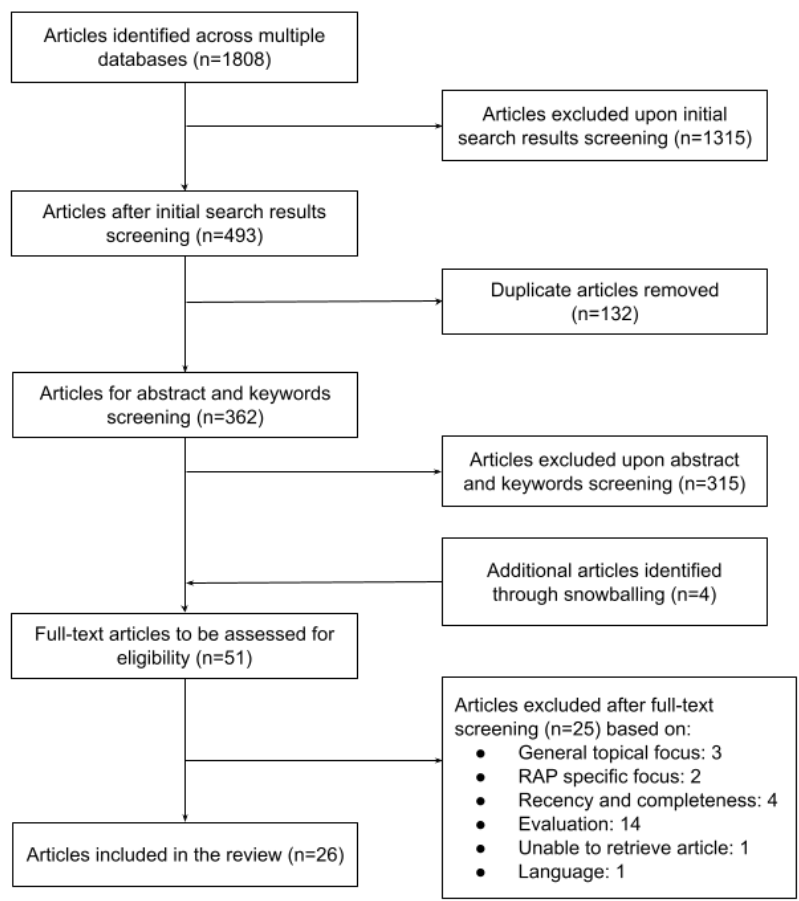

Figure 1: The flow diagram based on PRISMA-ScR [45] for the study selection process.

## 5.2 Results Synthesis

Table 4 provides some statistical information about our study sample, namely the number of articles per year, per country of the first author (based on the author's affiliation stated in the article), and per publication type.

To answer our research questions (RQs), we have extracted data from the selected articles, based on the data charting variables given in Table 3. The extracted data are summarised in two main tables, Table SDE_Q1-Q4 and Table SDE_Q5, which are, due to their size, made available in the "Data Extraction" supplementary (spreadsheet) file[7]. The data summarised in Table SDE_Q1-Q4 cover the first four groups of charting variables (Table 3) and are used to answer the first four RQs, whereas Table SDE_Q5 contains data about the evaluation of the RAP methods proposed in the selected studies and, thus, was used to answer RQ5.

In the following subsections, we address each RQ in turn, using the data from Tables SDE_Q1-Q4 and SDE_Q5, which have been aggregated according to the requirements of the corresponding RQ. An overview of the synthesised data for all research questions is given in Figure 2.

Table 4. Basic summary information about articles in the study sample

| | | |
|---|---|---|
| Year of publication | 2021 | 4 |
| | 2020 | 7 |
| | 2019 | 5 |
| | 2018 | 3 |
| | 2017 | 3 |
| | 2016 | 4 |
| Country of the 1st author | China | 13 |
| | USA | 3 |
| | India / Canada / Hungary | 2 each |
| | Italy / Poland / Spain / South Korea | 1 each |
| Publication type | Journal paper | 20 |
| | Conference paper | 6 |

---

[7] Table label SDE_Q1-Q4 stands for **S**upplementary **D**ata **E**xtraction for **Q**uestions 1-4. Similarly, SDE_Q5 stands for **S**upplementary **D**ata **E**xtraction for **Q**uestion 5. These labels are used to name the relevant sheets in the "Data Extraction" supplementary (spreadsheet) file available at: https://bit.ly/3KrJ7Ze.

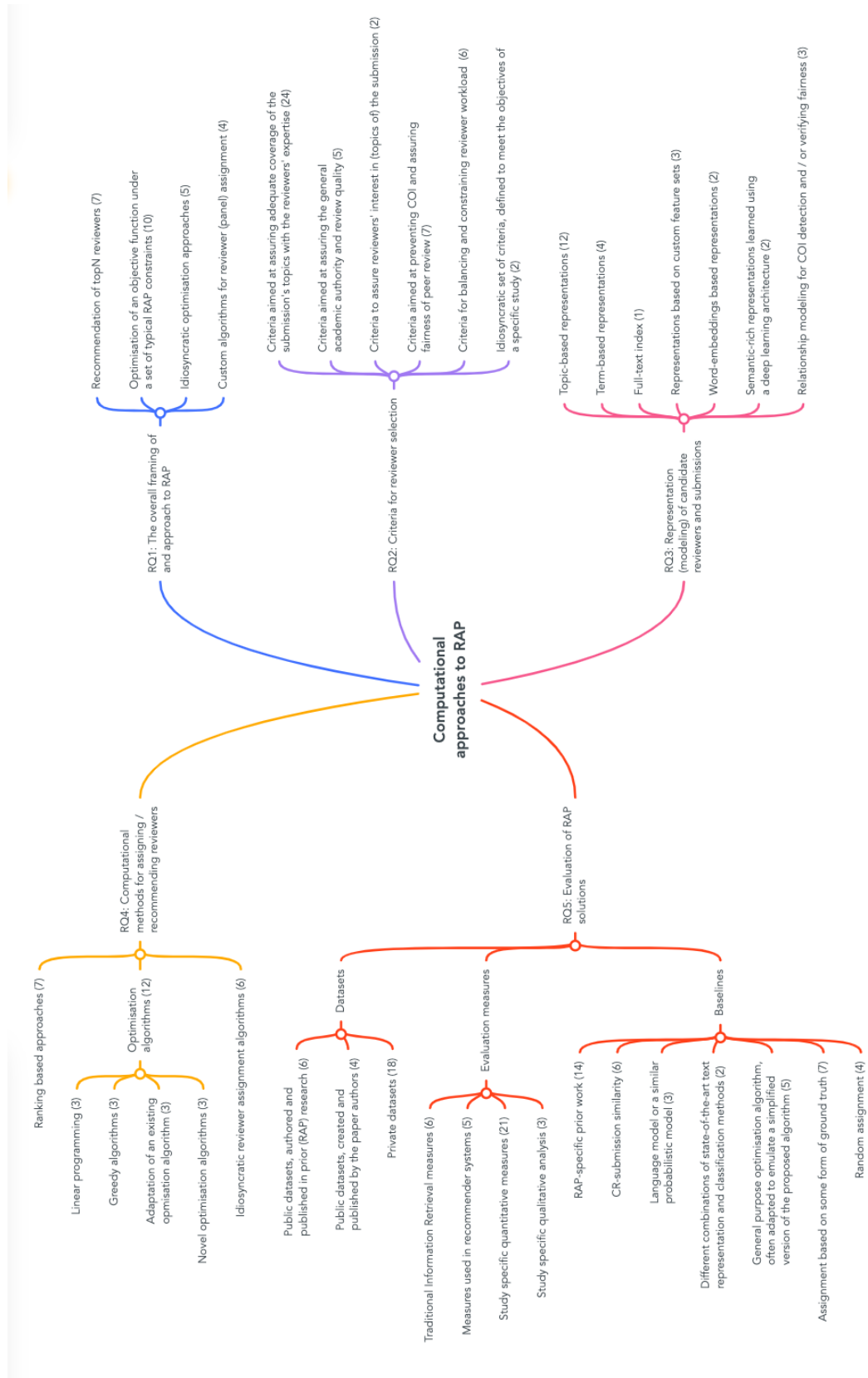

Figure 2: An overview of review findings in relation to each research question (RQ). Numbers in brackets in leaf nodes denote the number of papers that a certain approach / method applies to. The map was created using the MindMeister tool.

*5.2.1 RQ1: The overall approach to RAP*

Based on the framing of and the overall approach to RAP, we have identified, in the selected studies, four major groups of approaches. An overview of each group is given below, whereas a summary of the approach adopted in each individual study is given in Table 5.

*Recommendation of top N reviewers* (7/26 or 27% of studies). The ranking of reviewers is typically based on a custom defined, often topic-based, relevance of each candidate reviewer (CR) for a given submission [44, 1, 7, 12, 39]. However, there are also quite distinct approaches for the identification and ranking of reviewers such as the network-based approach by [37] or the weighted association rule mining approach by [3].

*Reviewer (panel) assignment through optimization of an objective function under a set of typical RAP constraints* (10/26 or 38% of studies). The assignment objective function is often defined in terms of the sum of matching scores of the submission-CR pairs. While the matching scores are computed in different ways, they always include, in one way or the other, topical relevance of a CR to the given submission (i.e., their topical similarity). Other elements that were considered for matching score computation include, for example, estimated quality of the CR's earlier reviews [34], collaboration distance [22] or CR's topical authority and indicators of research interest trends [13]. Typically considered constraints include conflict of interest (COI), workload per reviewer, and the number of reviewers per paper. In addition, some studies in this group include constraints such as local fairness (each submission is assigned to a reviewer panel that collectively possess sufficient expertise [22]) or absence of k-loops in assignments (loops are situations where the same persons act as both authors and reviewers and might be in a position to review each others' papers [9]).

*Reviewer assignment through an idiosyncratic optimisation approach* (5 / 26 or 19% of studies). Similar to the previous group, this one also includes studies that frame RAP as an optimisation task, but approach this task in a specific, idiosyncratic way; therefore, there is a large diversity in this set of proposed solutions. For example, Jecmen et al. [11] consider assignments as probabilistic and define the optimization objective as the expected sum of similarity scores across all submission-CR pairs, subject to the constraint on the probability of each assignment, as well as the typical load constraints. On the other hand, Pradhan et al. [36] define and solve RAP as an equilibrium multi-job problem; Liu et al. [23] propose a hybrid method that combines knowledge rules and an optimisation method, whereas Kalmukov [15] treats RAP as the task of maximising the sum of edge weights (which reflect the level of matching) of an undirected bipartite graph of CRs and submissions.

*Reviewer (panel) assignment based on a custom algorithm* (4 / 26 or 15% of studies). Similar to the previous group, the methods here are also very diverse as each study is specific in how it defines and/or approaches RAP. For example, Zhang et al. [55] framed RAP as a multi-class classification task and thus have developed a classifier that associated each submission with multiple topic labels from a predefined topic taxonomy. Based on the topic labels assigned to submissions by the classifier, the authors assign CRs to submissions based on the number of common topic labels. Zhao et al. [57] also start with classification, but they develop and use a classifier to associate each CR with research field labels from a predefined taxonomy. This is followed by assigning CRs to submissions based on the exact matching of research fields associated with CRs and submissions. Two studies [53, 31] employ an iterative assignment process to associate each submission with a panel of CRs in order to have a good coverage of the submission's topics within the panel. While focused on the same objective, the two studies employ very different iterative processes.

Table 5. Summary of the overall approaches to RAP in the analysed studies

| The overall framing of RAP | | Studies |
|---|---|---|
| Recommendation of topN reviewers | Ranking of candidate reviewers (CRs) using a multi-layered network approach that includes a topic network, a citation network, and a reviewer network; the first two networks provide input for the third, which is the core of the overall recommendation model. | 37 |
| | Ranking of CRs based on a custom defined, topic-based relevance of each CR for a given submission | 39, 44 |
| | A model of 'field knowledge' - built by applying sentence-pair modelling to title-abstract pairs of CRs' publications - is used to compute similarity of CR-submission pairs; the computed similarities are then used for CR ranking | 7 |
| | The proposed Author-Subject-Topic model provides input for the computation of CRs' posterior probabilites for the given submission; the computed probabilites are used for CR ranking. | 12 |
| | Ranking of CRs based on a score that reflects the similarity of CR and submission profiles to their shared topics. A topic is defined as a linear combination of word embeddings in CR and submission profiles. | 1 |
| | Weighted Association Rules (WARs) are mined and used for identifying strongly associated researchers, in order to: i) detect potential COIs in the current assignments (hence the need for additonal reviewers), ii) identify and rank additional (external) reviewers. | 3 |
| Reviewer (panel) assignment through optimisation of an objective function under a set of typical RAP constraints | The objecitve function defined as maximisation of CR's topical authority, research interests trend, and topical relevance | 13 |
| | The objective function defined as maximisation of the degree of topic-based CR-submission matching | 21, 52 |
| | The objective is to maximize the sum of CR-submission overall matching scores, where such matching scores are computed based on topic-based and term-based similarity of CR-submission pairs as weel as the estimated quality of the CRs' earlier reviews | 34 |
| | The objective is to simultaneously maximise topical relevance and minimise COI, where COI is not only explicitly given, but also inferred through analysis of academic networks (i.e., networks of researchers and of institutions). | 51 |
| | The objective is to simultaneously maximize the (topic-based) relevance of reviewers to the submission and the (topical) diversity of the reviewers as a group. | 30 |
| | The objective is the overall best assignment (i.e., maximising the sum of individual CR-submission matchings) while assuring *local fairness*, meaning that each submission is assigned to a reviewer panel that collectively possess sufficient expertise. | 17 |
| | The objective is to maximise the sum of individual CR-submission matching scores while assuring *k-loop free* assignments. Loops are situations where the same persons act as both authors and reviewers and might be in situation to review each others papers | 9 |
| | The objective is to maximise the sum of matching degrees (of CR-submission pairs), where matching degree is defined as a combination of collaboration distance as a fairness indicator and topic similarity as a measure of confidence. | 22 |
| | Optimisation task with the objective to simultaneously maximize the relevance of reviewers to the paper and the diversity of the reviewers as a group. | 30 |
| Reviewer assignment through an idiosyncratic | The objective is to maximise the sum of edge weights of an undirected bipartite graph of CRs and submissions (edge weights reflect the level of matching), subject to the constraint of balanced reviewer load and the requirement that each paper is assigned to at least one competent CR if such a CR is available | 15 |
| optimisation approach | The optimization objective is maximisation of the *expected* sum of similarity scores across all CR-submission pairs, subject to the constraint on the probability of each assignment, as well as the typical workload constraints. Note: assignments are probabilistic. | 11 |
| | Equilibrium multi-job problem where the objective is to find, for the given submission, an assignment scheme that for each CR maximizes topic similarity (of the CR-submission pair), minimizes COI, and balances the CRs' workload. | 36 |
| | A hybrid method that integrates knowledge rules and an optimisation method. The former assures absence of COI, while the latter maximizes the total discipline-based matching of CRs and submissions, subject to a set of typical RAP constraints. | 23 |
| | Iterative assignment process that in each step aims to simultaneously maximise the authority of selected CRs, the diversity of the reviewer panel, and the topical coverage of the submission (i.e., coverage of the submission's topics by the panel's expertise) | 53 |
| Reviewer (panel) assignment based on a custom algorithm | Iterative assignment process, where CRs are ranked based on a custom defined function of CR-submission matching and a set of typical RAP constraints, followed by diversity based selection of candidate reviewer panels | 10 |
| | Multi-label classification of submissions, which associates each submission with a subset of topic labels from a predefined topic taxonomy, followed by assigning CRs to a submission based on the number of shared topic labels with the submission. | 55 |
| | Classification of CRs, which associates each CR with research field labels from a predefined taxonomy, followed by research-field-based (exact) matching of CRs and submissions | 57 |
| | Iterative assignment process that uses an Ordered Weighted Averaging (OWA) aggregation function to combine several CR features (topical authority, recency, quality, availability) when ranking CRs for the given submission. The objective is to assign a group of CRs that offers a good coverage of the submission's topics. | 31 |

*5.2.2 RQ2: Criteria for reviewer assignment / recommendation*

A wide variety of criteria have been used for assigning reviewers to submissions. The full summary of the criteria identified in the included studies is given in Table 6.

The only criterion that is shared among a large number of studies (18/26, 69%) is the *topical relevance*, that is, topical similarity between a CR and the submission. This criterion, which essentially estimates how competent a CR is for the topic(s) of the submission, is computed based on the adopted topic-based representations of CRs and submissions (see Section 5.2.3).

Another criterion that is common to a subset of included studies is the *absence of conflict of interest* (COI), though it is primarily used as a constraint in studies that approach RAP as an optimization problem. In some studies [21, 9], information about COI is assumed to be given (i.e., directly available in the data), while in others it is inferred through analysis of academic networks [36, 51] or collected from a variety of sources and formalised as rules [23]. A closely related criterion is *fairness*, which is introduced in [22] to ensure equal treatment of all submissions by each reviewer (i.e., without being affected by direct or indirect COI) and that each submission is reviewed by the required number of distinct reviewers.

The *authority* of candidate reviewers is another criterion shared by a few studies. It refers to a CR's general academic standing and authority, and is in all cases, at least partially determined by the CR's h-index. In particular, it is either fully based on h-index [53], or is determined based on the CRs' citation counts and h-index [37], or computed as an aggregated score of several features, including the h-index, the number of PhDs supervised, and the number of papers published [31].

A variant of this criterion namely *topical authority* is used in two studies [13, 31] as an indicator of a CR's authority in submission-related topics.

Table 6. Summary of the criteria used for associating reviewers with submissions in the analysed studies

| Criteria for reviewer assignment / recommendation | | | Studies |
|---|---|---|---|
| Criteria aimed at assuring adequate coverage of the submission's topics with the reviewers' expertise | *Topical relevance,* topical similarity between a candidate reviewer (CR) and the submission (i.e., similarity of their topic-based representations) | | 1, 7, 12, 13, 14, 15, 22, 23, 30, 34, 36, 37, 39, 44, 51, 52, 55, 57 |
| | *Relevance*, defined as a linear combination of i) the *topic-based similarity* between the CR and the submission, and ii) the *referring value*, computed based on the common references (between a CR's publications and the submission) derived from co-citations network. | | 10 |
| | *Topical authority*, an indicator of a CR's recognition in submission-related topics | Estimated by applying an adapted version of PageRank on the researcher (CRs) citation network | 13 |
| | | Determined through (manual) mapping of the CR's (self-declared) topical interests and expertises to conference topics | 31 |
| | *Topical coverage*, a group level criterion, quantifies the extent to which all topics of the submission are covered by the expertise of a reviewer panel | | 53 |
| | *Affinity / suitability / similarity score* as a measure of matching between a CR and a submission. It is assumed that a matrix of CR-submission matching scores is given. While not explicitly stated, it is implied that these scores reflect topical matching between submissions and CRs. | | 9, 11, 17 |
| Criteria aimed at assuring the general academic authority and review quality | *Reviewer quality*, defined in terms of i) *quantity*, determined based on the number of publications, number of citations, and h-index; ii) *activity*, computed based on the number of publications, number of collaborators, and h-index in the past N years (N = 3..5) | | 10 |
| | *Review quality*, estimated based on: i) the difference between the CR's review score and the average of all the review scores, for each submission the CR reviewed, and ii) how timely, with respect to the deadline, the CR submitted their reviews. | | 34 |
| | *Recent research accomplishments*, operationalized as a weighted average impact factor of the CR's papers published in the past k years | | 31 |
| | *Authority,* general academic standing and authority | Determined based on the paper citation counts and h-index | 37 |
| | | Quantified based on the h-index | 53 |
| | | An aggregated score of several features, including the CR's h-index, the number of PhDs supervised, and papers published | 31 |
| Criteria to assure reviewers' interest in (topics of) the submission | *Recent research focus,* topics of recent publications are weighted more than older ones | | 37 |
| | *Interest trend*, specifically, direction and smoothness of the interest trend, as indicators of the CR's willingness to accept review invitation | | 13 |
| Criteria for balancing and | Limit on the number of assignments | | 9, 11, 21, 36 |
| constraining reviewer workload | | Balanced workload defined by the upper and lower bounds on the number of assignments | 17, 22 |
| Criteria aimed at preventing COI and assuring fairness of peer review | **Conflict of interest (COI)** | COI is defined as inversely proportional to co-authorship distance, which is determined from the co-autorship network | 36 |
| | | Inferred through analysis of academic networks; edge weights in these networks (as measures of strenght of the corresponding relationships) reflect the probability of COI | 51 |
| | | A variety of COI sources formalised as rules | 23 |
| | | Information about COI is assumed to be available (i.e., given) | 9, 21 |
| | *Diversity*, a group level criterion defined in terms of social influence, that is, the probability that one reviewer is influenced by another one | | 53 |
| | *Fairness*, a composite criterion that includes i) equal treatment of all submissions by each reviewer (i.e., without being affected by direct or indirect COI) and ii) having each submission reviewed by the required number of distinct reviewers | | 22 |
| Idiosyncratic set of criteria, defined to meet the objectives of a specific study | Criteria specific to finding additional reviewers for submissions with critical assignments | COI with the submission authors and the already assigned reviewers who do not need to be substituted | 3 |
| | | High association with the already assigned reviewer(s) with maximal confidence on the given submission | |
| | Criteria specific to the task of mitigating manipulation in peer review | Upper bound on the probability of a particular CR-submission pair | 11 |
| | | Upper bound on the probability of two specific CRs being assigned to the same submission | |

A small subset of studies [9, 17, 11] is focused exclusively on the assignment algorithm assuming that the extent of matching between submission-CR pairs is already available. More precisely, these studies assume that a matrix of submission-reviewer matching scores - the primary criterion for reviewer selection - is directly available. Such scores are referred to as affinity [17] or suitability [9] or similarity [11] scores and are used together with constraint criteria (e.g., absence of COI and/or reviewer workload) in the proposed algorithms.

Other identified criteria are unique to individual studies and not observed in more than one paper. They include, for example, a CR's *interest trend* (namely direction and smoothness of the trend), which is used as an indicator of the CR's willingness to accept a review invitation [13], or review quality, as an indicator of the *quality of reviews* previously completed by the CR [34]. In case of reviewer panel assignment, Yin et al. [53] introduced the *diversity* criterion, which they defined as the probability that one reviewer, within the panel, will be influenced by others. *Topical coverage* is another group level criterion, which quantifies the extent to which all topics of the submission are covered by the expertise of a particular reviewer panel [53].

*5.2.3 RQ3: Computational approaches to reviewer and submission modelling*

We have identified seven distinct groups of computational representations of CRs and submissions, which we describe below. A summary of the modelling approaches used in individual studies is given in Table 7.

Table 7. Summary of the computational approaches to reviewer and submission modelling in the analysed studies

| Reviewer and submission modelling / representation | | Studies |
|---|---|---|
| Topic-based representations | Topic vectors (weighted or unweighted), representing distributions over topics identified with the Latent Dirichlet Allocation (LDA) topic model | 10, 34, 36, 37, 51 |
| | Ordered sequences of topics identified using LDA; this ranking-based approach is used to reduce overfitting to incomplete researcher (CR) data. | 44 |
| | Distributions over topics identified by combining Author Topic Model (ATM) [42] with Expectation maximisation algorithm | 13 |
| | Topic vectors obtained from ATM | 3 |
| | Vectors of most relevant topics identified with LDA | 52 |
| | Distributions over topics identified with the Author-Subject-Topic model (topic model proposed by the paper authors) | 12 |
| | Distributions over topics obtained with Reviewer Aspect Model [16] | 53 |
| | Distributions over topics obtained through Latent Semantic Indexing (LSI) | 21 |
| Term-based representations | Vector space model (VSM) with TF-IDF weighting | 34, 37 |
| | Weighted term-space model, where weights account for the difference in importance of the same terms in different research areas | 30 |
| | Different variants of VSM are explored, including different ways of text preprocessing and different ways of computing term weights | 14 |
| Full-text index | A full-text index of a document that compiles the CR's research activities (reviews, projects, publications, ...) | 39 |
| Representations based on custom feature sets | CR: features quantifying topical authority (expertise in conference topics), recent research accomplishments, general research authority, and availability<br>Submission: vector of conference topics covered by the submission | 31 |
| | CR: a vector of keywords associated with weights that reflect the CR's expertise level and account for the change in the CR's interest over time and the frequency of publishing on a particular topic.<br>Submission: vector of keywords | 39 |
| | The (self-)declared levels of association between CRs / submissions and predefined disciplinary areas. Submissions are also assigned ranks based on an initial assessment of their quality | 23 |
| Word-embeddings based representations | Tags (keywords) encoded as word vectors, that is, represented in a vector (embeddings) space built using the Word2Vec skip-gram model | 57 |
| | CR: matrix with (Word2Vec) embeddings of all unique words in concatinated abstracts of the CR's publications.<br>Submission: matrix with (Word2Vec) embeddings of all unique words in the submission's abstract | 1 |
| Semantic-rich representations learned using a deep learning architecture | HIErarchical and transPArent Representation (HIEPAR) learnt with a deep neural network based on a two-level bidirectional gated recurrent unit and a two-level attention mechanism | 55 |
| | Sentence-pair modelling is applied to title-abstract pairs of CRs' publications to learn the CRs' "field knowledge"; this is done with a deep learning architecture that integrates an embeddings model (e.g., BERT) and a network-based learning algorithm (e.g., CNN) | 7 |

| Relationship modeling for COI detection and / or verification of the fairness criterion | Two kinds of undirected, weighted networks are constructed: networks of researchers and networks of institutions. Edge weights in these networks reflect the strenght of the corresponding relationships between researchers (CRs and authors) and thus the probability of COI. | 51 |
|---|---|---|
| | A weighted undirected network of researchers is constructed with nodes representing CRs and their weights reflecting the CRs' academic importance, whereas edge weights quantify the magnitude of collaboration between CRs. | 53 |
| | An unweighted undirected academic network is used for checking the fairness criterion. Nodes are CRs, authors of submissions, and other scholars in an academic digital library, while edges represents different kinds of academic relationships (co-author, co-affiliation, and advisor-advisee). | 22 |
| Reviewer - submission matching scores are either given or can be easily computed based on the available information. | It is assumed that submission-reviewer matching scores are given | 9, 11, 15, 17 |
| | It is assumed that topics (keywords) are associated with CRs and submissions, and that topic similarities are (pre)defined and available. | 22 |

*Topic-based representation is the most dominant approach in the reviewed literature* (12/26 or 46% of studies). This approach is often based on the Latent Dirichlet Allocation (LDA) [2] topic modelling method, in the context of which CRs and submissions are represented as (weighted or unweighted) topic vectors [34, 51, 36, 37, 10] or ordered sequences of the identified topics [44] or vectors of most relevant topics identified with LDA [52]. Other topic modelling methods that have been used in the selected studies include Author Topic Model [42], which was adopted in [13, 3]; Author-Subject-Topic model, which was proposed and used in [12]; Reviewer Aspect Model [16] used by Yin et al. [53], and Latent Semantic Indexing [20] used in [21].

*Term-based representation*, also known as vector space model (VSM). VSM is a widely used text representation method that models a piece of text as a set of terms (i.e., sequences of one or more consecutive words) it consists of. Each term is assigned a weight that quantifies its estimated relevance in the given text. An often used term-weighting method is Term Frequency - Inverse Document Frequency (TF-IDF), which weights terms based on the frequency of their occurrence in the given text (TF), but also their uniqueness at the level of the overall corpus (IDF). Two studies [37, 34] used VSM with the TF - IDF weighting scheme. Kalmukov [14] tested different variants of VSM, including different ways of text preprocessing and computing term weights. Finally, Mirzaei et al. [30] used a more sophisticated weighted term-space model, where weights account for the importance of the same terms in different research areas.

*Custom feature set representation*. Three studies used a custom defined set of features as the primary [23, 39] or the only [31] way of representing CRs and submissions. For example, to model CRs, Nguyen et al. [31] relied on a set of features quantifying the CR's topical authority, general research authority, recent research accomplishments, and availability, whereas Protasiewicz et al. [39] modelled CRs with vectors of keywords associated with weights that account for the change in a CR's interest and expertise over time.

*Full-text index*. This approach to CR and submission modelling was used by Protasiewicz et al. [39]. To model a CR, the authors built a full-text index of a document that is a compilation of the CR's research activities (reviews, projects, publications, and the like).

*Word-embeddings based representation*. Word embeddings are distributed word representations where words are mapped to dense vectors of real numbers [28]. By leveraging word context (i.e., surrounding words), word embeddings capture, to a certain extent, word meaning. Considering the wide adoption of word embeddings in various natural language processing (NLP) tasks, it was somewhat surprising that only two of the selected studies relied on this type of representation. In one of the studies, Anjum et al. [1] use pre-trained Word2Vec [28] embeddings to model CRs. In particular, each CR is represented as a matrix of word embeddings of all the unique words that appear in the concatenation of the abstracts of the CR's publications. Likewise, a submission is modelled as a matrix of word vectors of all unique words in the submission's abstract. In the other study, Zhao et al. [57] assume that both submissions and CRs are described with tags that reflect the keywords of submissions and the research interests of CRs. These tags are represented as vectors in an embedding space built using the Word2Vec skip-gram model [28].

*Semantic-rich representations learned using a deep learning architecture*. This group also includes two studies only. In one of them, Zhang et al. [55] propose HIErarchical and transPArent Representation (HIEPAR) as an advanced way of capturing and representing the semantics of submissions and CR profiles. This representation is learned using a complex attention-based deep learning architecture. The other study [7] applies sentence-pair modelling to title-abstract pairs of CRs' publications to learn the CRs' "field knowledge". This is done with a deep learning architecture that integrates an embeddings model (e.g., BERT) and a network-based learning algorithm (e.g., CNN). The small number of studies in this and the previous (word-embeddings) group suggests underrepresentation of state-of-the-art NLP approaches in RAP-focused research, which we discuss further in Section 6.

*Relationship modelling for COI detection and / or verifying fairness*. In addition to modelling the expertise of candidate reviewers, some of the studies model relationships among researchers and / or academic institutions to check for the presence of COI and to assure fair assignments. All of these studies [53, 22, 51] create undirected, weighted academic networks where nodes represent researchers / scholars and edges represent different kinds of relationships, all indicative of some form of (past and/or present) collaboration, and the strength of collaboration is reflected in the edge weights. Having constructed such networks, some researchers [22, 51] identify the presence of COI by computing the length of (weighted) shortest paths between the node corresponding to a CR and the nodes corresponding to each author of the given submission. If the shortest of the computed paths is below the given threshold, which can be either predefined [22] or set as a parameter [51], the considered CR is identified as having COI for the given paper. Yin et al. [53] leverage an academic (i.e., co-authorship) network to compute a measure of social influence that one researcher might have on another, and use that measure to assure fair reviewer assignment. They define social influence in terms of inverse social distance (i.e., the shortest path) between nodes and local connectivity of nodes (i.e., the number of direct collaborators).

*5.2.4 RQ4: Computational methods for assigning / recommending reviewers to submissions*

We have identified three groups of methods that have been used for assigning / recommending individual reviewers or reviewer panels to submissions, which we cover in the following. A summary of the reviewer assignment / recommendation methods reported in individual studies is given in Table 8.

*Ranking based approaches*. This group includes seven studies (27% of all selected studies), each one performing the ranking task based on a custom-defined candidate reviewer score. For example, in [7], the ranking score is the sum of similarities of the submission and k papers by CR that are the most similar to the given submission, whereas in [39], it is the total similarity (of a submission and a CR), defined as a linear combination of keywords-based similarity and full-text-index based similarity scores. A fairly unique approach is presented by Pradhan et al. [37] who applied Random Walk with Restart (RWR) to the Reviewer Network, which includes CRs as nodes, while edge weights are defined as a linear combination of several features representing diverse aspects of similarity of two CRs (including similarity of co-authorship profiles, h-index based similarity, similarity of research interests, similarity of publication venues, as well as direct citation and co-citation based similarity). Another unique approach is the one presented by Cagliero et al. [3] who tackle a specific version of RAP. In their work, the starting assumption is that the initial assignment of reviewers has already been done and the focus is on finding additional (external) reviewers for submissions with critical assignments (i.e., those that require additional reviews). With that backdrop, RAP is framed as a Weighted Association Rule (WAR) mining task and external reviewers are identified and ranked based on i) the absence of COI with the submission authors and internal reviewers[8] who do not need to be substituted, and (ii) high quality WARs that include the given external reviewer and internal reviewer(s) with maximal confidence (i.e., topical expertise) on the given paper[9].

*Optimisation algorithms.* This is the largest group, with 12 (46%) of the selected studies applying an optimisation algorithm to solve RAP. Three studies proposed some form of a greedy algorithm [30, 34, 53]. For example, to determine an optimal assignment of reviewer panels to a given set of submissions, Mirzaei et al. [30] proposed a greedy forward-selection algorithm that simultaneously maximises the sum of the submission-CR relevance scores of all CRs within a group and minimises the sum of pairwise similarities of the CRs in the group; the algorithm also prioritizes submissions for which only few relevant CRs are available in the reviewer pool. Another three studies [13, 9, 21] used some form of linear programming to optimise a (custom) defined objective function under the typical set of RAP constraints. Among these three studies, the one by Guo et al. [9] distinguishes itself as it proposes algorithms that combine linear programming and graph-based search to ensure optimal k-loop free assignments[10]. The other six studies in this group proposed either novel optimisation algorithms [11, 17, 22] or adaptations of existing ones to their formulation of RAP [52, 36, 51].

*Idiosyncratic assignment algorithms.* This group includes six studies with assignment algorithms that are not only quite distinct from algorithms proposed in the other two groups, but are also mutually distinct. For example, Zhao et al. [57] compute similarities between CRs and submissions (i.e., their Word2Vec embeddings) with the Word Mover Distance [19] method. The computed similarities serve as the input to the Constructive Covering Algorithm to build a classification model that predicts research interests of CRs, using labels of submissions as supervisory information. As a result, CRs are associated with (predicted) research labels and are, then, assigned to the submissions based on the matching of research field labels. Also interesting is the two-phase algorithm proposed by Hoang et al. [10] for assigning a reviewer group to a submission: the first phase of this algorithm includes ranking and selection of CRs for each submission, based on a custom-defined measure of CR-submission matching and subject to the typical RAP constraints; in the second phase, groups of eligible reviewers are composed and selected by considering COI and the group's diversity score (the latter is estimated based on the pairwise similarity of reviewers in the group). We will also mention that the method proposed by Liu et al. [23] is the only hybrid method among the selected studies, which combines optimization with knowledge-based rules. In particular, knowledge rules are first executed to eliminate potential COI; this is followed by the optimisation task, which maximises the total discipline-based matching of CRs and submissions under a typical set of RAP constraints.

---

[8] Internal reviewers are those who were initially assigned to a paper.
[9] WAR represents an implication between two sets of authors. Such rules are used for discovering significant co-author relationships.
[10] Loops are defined as situations where the same persons act as both authors and reviewers and might need to review each other's papers. Loops are detected on the assignments graph, where directed edges represent assignments from reviewer to author nodes. Loops of length > k are considered acceptable, while smaller ones are not.

Table 8. Summary of the computational to reviewer and submission modelling in the analysed studies

| | | Computational methods for assigning / recommending reviewers to submissions | Studies |
|---|---|---|---|
| Optimisation algorithms | Linear programming | Integer linear programing is used to maximise the objective f. defined in terms of topical authority, indicators of research interest trend, and topical relevance of CRs, subject to typical RAP constraints | 13 |
| | | Standard linear programing solver that maximizes the total suitability score (i.e., a measure of CR-submission topic-based matching) for all submissions | 21 |
| | | Two algorithms are proposed; both combine linear programing and graph-based search to ensure optimal *k-loop free* assignments | 9 |
| | Greedy algorithms | A greedy forward-selection algorithm that simultaneously maximises the sum of the submission-CR relevance scores of all CRs in a group and minimises the sum of pairwise similarities of the CRs in the group; the algorithm prioritizes submissions with fewer relevant CRs | 30 |
| | | A greedy algorithm that maximizes the objective f. defined in terms of i) the prior probability of a set of CRs acting as a team with well balanced authority and diversity, and ii) the probability that the team's topical coverage provides the expertise required by the given submission | 53 |
| | | A greedy algorithm that maximizes the total matching of CRs and submissions, where CR-submission matching score is a linear combination of topic-based and term-based similarity scores and a score of CR's review quality | 34 |
| | Adaptations of existing algorithms | A variant of the Bare Bones Particle Swarm Optimization algorithm is proposed and used to maximise topic-based objective f. (proportion of CR's relevant topics shared with the submissions) subject to typical RAP constraints | 52 |
| | | A meta-heuristic greedy algorithm is proposed to solve RAP as an equilibrium multi-job problem, that is, to find optimal job (=submission) assignment schedule where for each worker (=CR), the difference between max and min total cost (=combination of topical similarity, COI, and workload) is minimal. | 36 |
| | | The optimisation problem is transformed into an equivalent minimum convex cost flow problem and solved by constructing a directed network with a node for each submission and each CR and a convex cost fuction associated with each edge (connecting a submission and a CR) | 51 |
| | Novel algorithms | Two algorithms are proposed for solving the optimisation problem with the local fairness constraint (= each submission is assigned to a reviewer panel that collectively possess the required expertise): FairIr (FAIR matching via Iterative Relaxtion) and FairFlow | 17 |
| | | Two optimisation algorithms are proposed to simultaneously maximise total collaboration distance (as a fairness indicator) and total topic similarity (as a measure of confidence) subject to load balancing and absence of COI constraints | 22 |
| | | Two phase algorithm is proposed: 1st phase: an optimal "fractional assignment" matrix is built, with marginal probabilities of individual reviewer-paper assignments; 2nd phase: this matrix is used for assignment sampling while respecting the requirment not to assign two reviewers from the same subset (e.g., same institution) to the same paper. | 11 |
| Ranking based approaches | | Iterative ranking process based on i) the estimated relevance of a CR to the submission, and ii) the estimated relevance of the CR's papers to the submission; both relevance measures are defined in terms of topic-based and word-based similarity measures | 44 |
| | | Ranking of CRs based on the sum of similarities of the submission and CR's *k* papers that are most similar to the submission | 7 |
| | | Ranking of CRs based on the posterior probability of each CR for the given submission | 12 |
| | | Ranking of CRs based on a harmonic mean of reviewer-topic and submission-topic relevance scores, where the focus is on *common topics* (i.e., semantically similar topics) from CR and submission profiles | 1 |

| | | |
|---|---|---|
| | Ranking of CRs based on the CR-submission total similarity score defined as a linear combination of keywords-based similarity and full-text-index based similarity | 39 |
| | External reviewers are identified and ranked based on i) the absence of COI with the submission authors and internal reviewers who do not need to be substituted, and (ii) high quality WARs that include the given external reviewer and internal reviewer(s) with maximum confidence on the given paper. | 3 |
| | Random Walk with Restart (RWR) applied to the Reviewer Network; the network includes CRs as nodes, while edge weights are defined as a linear combination of several features representing diverse aspects of similarity of two CRs. Through an iterative process, RWR ranks all CRs in the network. | 37 |
| Idiosyncratic assignment algorithms | Two phase algorithm: 1st phase: ranking and selection of CRs for each submission, based on a custom defined measure of CR-submission matching and subject to the typical RAP constraints. 2nd phase: groups of eligible reviewers are composed and selected by considering COI and the group's diversity score. | 10 |
| | Multi-label based reviewer assignment (MLBRA) approach: after a submission is associated with top-k relevant topic labels, predicted by the proposed multi-label classification model, it is assigned to the CR who has the highest level of topical overlap with the given submission (i.e., the highest number of shared topic labels) | 55 |
| | Word Mover Distance based similarities between CRs and submissions (i.e., their embeddings) serve as the input for building a classification model that predicts research field labels of CRs; CRs whose predicted research fields match those of the submission get selected. | 57 |
| | An iterative algorithm where each iteration includes: i) selection of the submission with the highest (topical) coverage requirements, ii) ranking of CRs based on the estimated level of relevance to the submission, iii) assigning the top ranked CR and updating the submission's coverage requirements and the CR's availability | 31 |
| | The algorithm starts with the similarity matrix and for each submission, identifies the most similar (i.e., competent) CR. If that results in higher work load for some CRs, similarity scores are altered to give priority to submisions with a small number of CRs competent to evaluate them. Each iteration of the algorithm assigns one reviewer to each submission. | 15 |
| | A hybrid method. First, knowledge-based rules are executed to eliminate potential COI; then, typical RAP optimisation is done: maximizing the total discipline-based matching of CRs and submissions under a set of constraints | 23 |

*5.2.5 RQ5: Evaluation of RAP solutions*

A summary of the key elements of the evaluation methods used in the selected studies is given in Table SDE_Q5[11]. For each study, the table includes a description of the overall evaluation method, the dataset(s) used, the baseline(s) used for comparison purposes, the evaluation measure(s), and the availability of the code that implements the proposed approach and/or the code used in the evaluation study. In the following, we provide a summary of the evaluation datasets, baselines, and measures from all the selected studies.

*Datasets*. Overall, there is a very small number of publicly available RAP datasets and true benchmarking datasets are not available. Furthermore, those few public datasets have been sparingly used. In particular, only six (out of 26) studies used one of the two existing public datasets, both created in earlier RAP-focused research:

---

[11] SDE_Q5 (**S**upplementary **D**ata **E**xtraction for **Q**uestion 5) sheet of the aforementioned "Data Extraction" supplementary file (https://bit.ly/3KrJ7Ze)

- The NIPS dataset by [29] contains 148 papers accepted for the 2006 edition of the Neural Information Processing Systems conference (NeurIPS 2006), a reviewer list of 364 members, and ground truth relevance judgments (for paper - reviewer pairs) made by prominent researchers from the NeurIPS community.
- The SIGIR dataset [16] includes 73 papers and 189 reviewers; the papers and the reviewers are manually annotated with ground-truth topics, which originate from the topics listed in the CfP of the SIGIR conference.

Three of the selected studies contributed new public datasets. Duan and colleagues [7] published two ground truth datasets, though the ground truth was based on heuristic rules[12]. These datasets were used by the same group of researchers in their more recent study [44]. Other public datasets were published by Kobren et al. [17] and Jecmen et al. [11] and in both cases, the published data are in the form of submission-reviewer similarity / affinity scores.

The majority of the selected studies built and used their own datasets, but did not make them publicly available. Among these, eight studies created and used some form of ground truth datasets, whereas the datasets built and used in the other ten studies did not include ground truth but were used to compare the proposed method against the selected baselines.

A summary of the datasets used across the selected studies is given in Table *Datasets*, which is, together with other summary tables for RQ5, made available in the "RQ5 Data Summary" supplementary (spreadsheet) file[13], in order to adhere to the page limit.

*Baselines*. Table *Baselines* in the supplementary file[14] summarises the baselines used across the selected studies. The absence of true benchmarks is also evident in the selection of methods used as baselines. A wide variety of baselines were used, often RAP methods reported in the literature (14 out of 26 studies), among which, those proposed in [29, 4, 18] were the most frequent ones. In six studies, similarities between candidate reviewers and submissions, based on different representational methods (e.g., topic-based, term-based, word embeddings) and different similarity measures (e.g., cosine, word mover distance), were used as the baselines. Seven studies relied on comparisons with some form of ground truth as either the only or one of the ways of assessing the validity of the proposed approach. Other kinds of baselines include language models or similar probabilistic models (3 studies); a combination of state-of-the-art text representation and classification methods (2); general optimisation algorithms, often slightly adapted to emulate a simplified version of the proposed algorithm (5); and random assignment (4).

*Evaluation measures*. A summary of the evaluation measures used across the selected studies is given in Table *EvalMeasures* in the supplementary file[15]. Only a small subset of studies (7 out of 26) used traditional Information retrieval (IR) evaluation measures (e.g., precision@k, recall@k, F1, mean averaged precision) or measures typical for evaluation of recommender systems (e.g., normalized discounted cumulative gain - nDCG). Even these studies tended to combine IR measures with custom-defined evaluation measures. For example, Pradhan and colleagues [37] used *Precision@k* and *nDCG* together with four custom-defined and RAP-specific measures including authority of *top-k reviewers*, *expertise matching score*, *diversity of reviewers*, and *reviewer coverage*. All other studies (19 out of 26) relied exclusively on a variety of evaluation measures that were tailored to the particularities of the proposed RAP approach. For example, studies that were aimed at assigning a reviewer panel to each submission typically combined measures of topical coverage of the submission by the panel's expertise with a measure of the panel's diversity. As an illustration of this practice, Mirzaei et al. [30] combined *Average confidence* as a measure of redundancy in reviewers' topical expertise, with the *Paper topical coverage* measure, both originally proposed in Karimzadehgan et al. [16]. A somewhat different set of measures were used by Yin et al. [53] who evaluated their proposal using the measures of *Authority* (normalized sum of h-index values of the reviewer panel), *Diversity* (inverse of the social influence within the reviewer panel), *Coverage* (Paper topical coverage measure, by Karimzadehgan et al. [16]), and *F-score* (harmonic mean of Diversity and Coverage). A small number of studies applied some form of manual qualitative analysis of assignments [52, 23] or reviewers' perceptions [21], to get additional insight into the quality of the produced assignments.

---

[12] For example, a researcher who has published at least 10 papers in a field corresponding to a manuscript is considered eligible to review that manuscript.
[13] Datasets sheet of the "RQ5 Data Summary" supplementary file (https://bit.ly/3kGfnNG)
[14] Baselines sheet of the "RQ5 Data Summary" supplementary file (https://bit.ly/3kGfnNG)
[15] EvalMeasures sheet of the "RQ5 Data Summary" supplementary file (https://bit.ly/3kGfnNG)

## 6 DISCUSSION

In this section, we, first, discuss the study results in relation to each research question, then lay out the study limitations, and conclude with an outline of possible future research directions.

### 6.1 Summary of Evidence

***Our first research question*** was focused on the overall approach to RAP, that is, how RAP was framed and what the overall method to solving this problem was. We have observed a lot of diversity among the selected studies, which made it quite difficult to identify distinct groups of approaches. Eventually, we identified four broad groups: *recommendation of top N reviewers*, reviewer (panel) *assignment through optimization of an objective function under a set of typical RAP constraints*, reviewer *assignment through an idiosyncratic optimisation approach*, and reviewer (panel) *assignment based on a custom algorithm*. Within each of these groups, there is a lot of heterogeneity, as some studies distinguish themselves by the particularities of their focus whereas others by their idiosyncratic methodology. In the former group are, for example, studies by Jecmen et al. [11], Cagliero et al. [3], and Guo et al. [9], where the focus is on ensuring fairness of the review process that goes beyond controlling for explicitly stated COI. In particular, aiming to ensure fairness, Jecmen et al. [11] focus on overcoming the practises of untruthful favourable reviews and torpedo reviewing; Cagliero et al. [3] focus on finding additional reviewers for submissions with critical assignments (i.e., assignments that are assessed as not fully adequate); whereas Guo et al. [9] aim at preventing assignments that correspond to small loops in a bipartite network of authors and reviewers, so that situations where two scholars (who are both authors and reviewers) review each other's papers are avoided. The group of studies with idiosyncratic methodology include, for example, Duan et al. [7] who applied a sentence pair modelling approach to title-abstract pairs from CRs' publications to model the CRs' 'field of knowledge'; Anjum et al. [1] who used pre-trained word embeddings and abstract topic vectors to derive and match semantic representations of the submission and the expertise of CRs; or Pradhan et al. [37] who proposed a multi-layered network approach, where each network provided input to the next one, and ranking of CRs was done by applying the Random Walk with Restart algorithm on the last network in the sequence (i.e., the reviewer network).

*The second research question* was about the criteria used for selecting 'eligible' reviewers. While we were able to identify a range of criteria in the literature, there is only one that stands out by the frequency of its use - the *topical relevance* criterion, which is often computed as the similarity between topic-based representations of a CR and a submission. *Absence of conflict of interest* and *reviewer authority* are two additional criteria that have been used across studies, but far less often than topical relevance. The other criteria appeared mostly in a single study. Still, it should be noted that some criteria reflect the same or very similar intention, but have been operationalized in different ways. For example, *recent research focus* in [37] and *interest trend* in [13] reflect the same underlying idea that higher relevance should be given to CRs' recent research topics, but the two have been operationalized in different ways. Some of the criteria that were sparingly used probably deserve more attention. For example, only one study [34] considered the quality of the CR's past reviews. Considering the frequently reported issues with review quality [27], this criterion requires more attention. Furthermore, since it is often not easy to find experts who are willing to accept review requests [49, 46], it would be important to put more emphasis on the reviewer interests or other similar criteria that may result in assignments that would be intrinsically motivating, thus increasing the likelihood of accepting the review request. *Availability* is another criterion that has received little attention in studies focused on recommendation / assignment of individual reviewers (i.e., journal context). Given the importance of this criteria when selecting reviewers for a journal submission[16], future RAP solutions could consider it as one of their constraints. In contrast, reviewer availability has been well covered in RAP solutions oriented towards conference-like settings, as several studies included reviewer workload (i.e., limits on the number of assignments) as one of the constraints. Last but not the least important, considering the relevance of fairness in peer review and the fact that is not always at the desired level [43], it would be important to give more attention to *fairness* as a multifaceted reviewer selection criterion that extends beyond directly observable and/or self-reported COI, to also consider potential indirect connections between authors and CRs (e.g., identified through academic networks), equal number of reviewers per submission, as well as adequate submission coverage in terms of complementarity of the reviewers' expertise.

Through **the third research question**, we have explored modelling approaches that were used for computational representation of CRs and submissions. Regarding CRs, the primary and often the only subject of modelling was their expertise, as evidenced in their published work. Only a handful of studies modelled relationships among scholars and/or scholarly institutions for the purpose of COI detection [51], assessing fairness [22], or ensuring diversity within a reviewer panel [53]. In the majority of included studies, modelling of CRs' expertise and the submission content was based on traditional NLP techniques, primarily topic models, vector space model, and keywords. Only a small subset of studies applied state of the art NLP methods for generating semantic rich representations, such as word embeddings [57, 1] or semantic representations learned through deep neural network models [55, 7]. This could be an area for further growth given the large interest in and extent of use of such models in a variety of NLP tasks and domains [24, 41, 32]. As a final note, while limited in quantity, more recent representational models identified in the selected studies can be considered as notable advancements, both in terms of novelty and diversity, over earlier RAP approaches that were typically limited to keywords (free-form or from controlled vocabularies) or some variant of vector space model [38].

To answer **the fourth research question**, we explored computational methods that have been used for selecting (assigning / recommending) individual reviewers or reviewer panels for submissions. This led to the identification of three broad groups of methods, namely ranking based approaches, optimisation algorithms, and idiosyncratic assignment algorithms - each with a common trait shared by the included algorithms, but also a lot of diversity and idiosyncrasy within each group (as shown in Section 5.2.4). Custom made or customised optimization algorithms constitute the largest group of computational methods used for reviewer assignment. This was expected considering the long tradition of optimization algorithms in RAP-related research [47]. Still, many algorithms from this group are quite innovative (e.g., [17, 30, 11]). Likewise, a fair number of studies in the other two groups offered novel reviewer assignment / recommendation methods (e.g., [3, 44, 55, 57]), thus opening many directions for future RAP-related research.

The objective of **our fifth research question** was to explore how RAP solutions have been evaluated and what, if any, benchmarks have been used. Regarding the datasets used for evaluation purposes, a large majority of the selected studies relied on self-curated datasets that were typically built by collecting data about papers and authors from academic digital libraries such as DBLP, ACM Digital Library[17], and Wanfang[18]; repositories such as AMiner[19] or arXiv[20]; or conference proceedings (e.g., SIGKDD, ICL, ICBIM). However, only a small number of studies that built their own evaluation datasets made those datasets publicly available. It should be also noted that there are no true benchmarking RAP datasets and only a small subset of studies used one of the two publicly available datasets that were published in earlier RAP-focused research work [29, 16]. We note that the two available datasets are more than 15 years old. The heterogeneity is also evident in the used baselines. It can probably be attributed, again, to the absence of benchmarks, but also to the lack of publicly available implementations of the RAP approaches published in the literature, which makes it more difficult to reproduce the RAP methods reported in earlier studies. The used evaluation measures are also quite distinct and numerous. This finding is consistent with the one made by Patil and Mahalle [33] in their review of performance measures used for the evaluation of RAP solutions. In addition to several traditional information retrieval evaluation measures, the selected studies used a myriad of custom defined evaluation measures, many of which were used in one study only. Regarding replicability, only two of the selected studies [17, 11] made the full code and evaluation datasets publicly available on GitHub, whereas one study [15] provided detailed pseudocode via an appendix.

---

[16] Many journals have a policy that a reviewer should not be invited if they are currently assigned to another paper or have recently completed a review.
[17] https://dl.acm.org/
[18] http://www.wanfangdata.com/
[19] https://www.aminer.org/
[20] https://arxiv.org/

## 6.2 Limitations

We note that one of the limitations of the current study can be related to the used search terms. By studying the literature, prior to initiating this study, we identified the terms that had been used to refer to the task of assigning / recommending reviewers to submissions in typical academic peer review contexts. We used those terms to formulate our search queries (Section 4.1). However, it might be the case that some authors used different terminology to refer to the same task, in which case, we might have missed their work. We consider this unlikely, since none of the studies we identified through additional search (i.e., through snowballing, Figure 1) differed in the terminology from the articles collected through the initial search. The second study limitation can be related to the fact that we considered only studies that were published in English. While this is a usual practice, in our case it might be limiting since half of the selected studies are by Chinese authors (13 out of 26), suggesting that there might be relevant RAP-related research that was published in Chinese. Finally, we note that we had to exclude one article that we did not have access to.

## 6.3 Recommendations for future research

Our findings point to several directions for future RAP-related research:

- First, even though a notable portion of studies framed RAP as a recommendation task, none of them examined the potentials of state-of-the-art recommendation algorithms, such as those based on deep learning models [56]. Hence, a research direction worth exploring is how recommendation models can be applied and / or adapted to RAP.
- Second, modelling of candidate reviewers and submissions has been predominantly based on traditional NLP methods and techniques (see Section 5.2.3). However, NLP community has offered advanced language models and architectures, such as different kinds of word / sentence / document embeddings [54, 41], which may be used in RAP to advance the semantic representation of candidate reviewers' expertise and the content of submissions, eventually enabling better matching of the two.
- Third, while a few of the selected articles proposed full decision support systems for reviewer selection, none anticipated the presence of an explanation facility. Considering the increasing complexity of the algorithms used for solving RAP and the general trend towards higher algorithmic transparency [48], it is advisable that future RAP-related methods include a component for offering explanations of the suggested assignments / recommendations.

- Fourth, related to the evaluation of the proposed RAP solutions, there is a clear need for a public benchmark with well-defined procedure, dataset(s), and evaluation measures. The creation of such a benchmark is challenging since RAP is not a well-defined task, as evident in the variety of ways it was framed in the selected studies. The complexity in creating such a benchmark is also evident in the fact that it was recognized as needed more than a decade ago [47] and still not available. Still, it is a worthy effort, considering that the availability of such benchmarks contributed to the progress in solving other complex problems (e.g., progress on a variety of information retrieval tasks has been incited and supported by benchmarks offered for the TREC[21] series of workshops).
- Last but not the least, we would highlight the relevance of making code and data publicly available to facilitate replication and benchmarking, and at the same time contribute to the open science initiative[22].

---

[21] https://trec.nist.gov/overview.html
[22] https://www.cos.io/

Survey. *arXiv:2003.08271 [cs]* (June 2021). DOI:https://doi.org/10.1007/s11431-020-1647-3

[42] Michal Rosen-Zvi, Thomas Griffiths, Mark Steyvers, and Padhraic Smyth. 2004. The author-topic model for authors and documents. In *Proceedings of the 20th conference on Uncertainty in artificial intelligence* (UAI '04), AUAI Press, Arlington, Virginia, USA, 487–494.

[43] Nihar B. Shah and Zachary Lipton. 2020. SIGMOD 2020 Tutorial on Fairness and Bias in Peer Review and Other Sociotechnical Intelligent Systems. In *Proceedings of the 2020 ACM SIGMOD International Conference on Management of Data* (SIGMOD '20), Association for Computing Machinery, New York, NY, USA, 2637–2640. DOI:https://doi.org/10.1145/3318464.3383129

[44] Shicheng Tan, Zhen Duan, Shu Zhao, Jie Chen, and Yanping Zhang. 2021. Improved reviewer assignment based on both word and semantic features. *Inf Retrieval J* 24, 3 (June 2021), 175–204. DOI:https://doi.org/10.1007/s10791-021-09390-8

[45] Andrea C. Tricco, Erin Lillie, Wasifa Zarin, Kelly K. O'Brien, Heather Colquhoun, Danielle Levac, David Moher, Micah D.J. Peters, Tanya Horsley, Laura Weeks, Susanne Hempel, Elie A. Akl, Christine Chang, Jessie McGowan, Lesley Stewart, Lisa Hartling, Adrian Aldcroft, Michael G. Wilson, Chantelle Garritty, Simon Lewin, Christina M. Godfrey, Marilyn T. Macdonald, Etienne V. Langlois, Karla Soares-Weiser, Jo Moriarty, Tammy Clifford, Özge Tunçalp, and Sharon E. Straus. 2018. PRISMA Extension for Scoping Reviews (PRISMA-ScR): Checklist and Explanation. *Ann Intern Med* 169, 7 (October 2018), 467–473. DOI:https://doi.org/10.7326/M18-0850

[46] Inga Vesper. 2018. Peer reviewers unmasked: largest global survey reveals trends. *Nature* (September 2018). DOI:https://doi.org/10.1038/d41586-018-06602-y

[47] Fan Wang, Ning Shi, and Ben Chen. 2010. A comprehensive survey of the reviewer assignment problem. *Int. J. Info. Tech. Dec. Mak.* 09, 04 (July 2010), 645–668. DOI:https://doi.org/10.1142/S0219622010003993

[48] Hugh Watson and Conner Nations. 2019. Addressing the Growing Need for Algorithmic Transparency. *Communications of the Association for Information Systems* 45, 1 (December 2019). DOI:https://doi.org/10.17705/1CAIS.04526

[49] Michael Willis. 2016. Why do peer reviewers decline to review manuscripts? A study of reviewer invitation responses. *Learned Publishing* 29, 1 (2016), 5–7. DOI:https://doi.org/10.1002/leap.1006

[50] Thomas Wolf, Lysandre Debut, Victor Sanh, Julien Chaumond, Clement Delangue, Anthony Moi, Pierric Cistac, Tim Rault, Remi Louf, Morgan Funtowicz, Joe Davison, Sam Shleifer, Patrick von Platen, Clara Ma, Yacine Jernite, Julien Plu, Canwen Xu, Teven Le Scao, Sylvain Gugger, Mariama Drame, Quentin Lhoest, and Alexander Rush. 2020. Transformers: State-of-the-Art Natural Language Processing. In *Proceedings of the 2020 Conference on Empirical Methods in Natural Language Processing: System Demonstrations*, Association for Computational Linguistics, Online, 38–45. DOI:https://doi.org/10.18653/v1/2020.emnlp-demos.6

[51] Sixing Yan, Jian Jin, Qian Geng, Yue Zhao, and Xirui Huang. 2017. Utilizing Academic-Network-Based Conflict of Interests for Paper Reviewer Assignment. *IJKE* (2017), 65–73. DOI:https://doi.org/10.18178/ijke.2017.3.2.089

[52] Chen Yang, Tingting Liu, Wenjie Yi, Xiaohong Chen, and Ben Niu. 2020. Identifying expertise through semantic modeling: A modified BBPSO algorithm for the reviewer assignment problem. *Applied Soft Computing* 94, (September 2020), 106483. DOI:https://doi.org/10.1016/j.asoc.2020.106483

[53] Hongzhi Yin, Bin Cui, Hua Lu, and Lei Zhao. 2016. Expert team finding for review assignment. In *2016 Conference on Technologies and Applications of Artificial Intelligence (TAAI)*, 1–8. DOI:https://doi.org/10.1109/TAAI.2016.7932314

[54] Tom Young, Devamanyu Hazarika, Soujanya Poria, and Erik Cambria. 2018. Recent Trends in Deep Learning Based Natural Language Processing. *arXiv:1708.02709 [cs]* (November 2018). Retrieved April 27, 2022 from http://arxiv.org/abs/1708.02709

[55] Dong Zhang, Shu Zhao, Zhen Duan, Jie Chen, Yanping Zhang, and Jie Tang. 2020. A Multi-Label Classification Method Using a Hierarchical and Transparent Representation for Paper-Reviewer Recommendation. *ACM Trans. Inf. Syst.* 38, 1 (February 2020), 5:1-5:20. DOI:https://doi.org/10.1145/3361719

[56] Shuai Zhang, Lina Yao, Aixin Sun, and Yi Tay. 2019. Deep Learning Based Recommender System: A Survey and New Perspectives. *ACM Comput. Surv.* 52, 1 (February 2019), 5:1-5:38. DOI:https://doi.org/10.1145/3285029

[57] Shu Zhao, Dong Zhang, Zhen Duan, Jie Chen, Yan-ping Zhang, and Jie Tang. 2018. A novel classification method for paper-reviewer recommendation. *Scientometrics* 115, 3 (June 2018), 1293–1313. DOI:https://doi.org/10.1007/s11192-018-2726-6
27